\documentclass[
    aip,
    jcp,
    preprint,
    amsmath,amssymb
]{revtex4-1}

\usepackage{amsmath}
\usepackage{amssymb}
\usepackage{graphicx}
\usepackage{dcolumn}
\usepackage{bm}

\usepackage[utf8]{inputenc}
\usepackage[T1]{fontenc}
\usepackage{mathptmx}
\usepackage{etoolbox}

\usepackage[ruled, lined, linesnumbered]{algorithm2e}

\newcommand{\rt}[1]{\textcolor{black}{#1}} 
\usepackage{soul}
\usepackage[normalem]{ulem}
\newcommand{\rsmath}[1]{\bgroup\markoverwith{\textcolor{red}{\rule[0.5ex]{2pt}{0.4pt}}}\ULon {\textcolor{red}{#1}}} 

\usepackage{eso-pic}
\AddToShipoutPictureBG*{%
  \AtPageUpperLeft{%
    \hspace{\paperwidth}%
    \raisebox{-\baselineskip}{%
}}}%

\makeatletter
\def\@email#1#2{%
 \endgroup
 \patchcmd{\titleblock@produce}
  {\frontmatter@RRAPformat}
  {\frontmatter@RRAPformat{\produce@RRAP{*#1\href{mailto:#2}{#2}}}\frontmatter@RRAPformat}
  {}{}
}%
\makeatother

\begin{document}

\preprint{APS/123-QED}

\title{A bond swap algorithm for simulating dynamically crosslinked polymers} 



\author{Peilin Rao}
 \thanks{These authors contributed equally.}
 \affiliation{School of Chemistry, Chemical Engineering and Biotechnology, Nanyang Technological University, 62 Nanyang Drive, Singapore 637459}
\author{Xiuyang Xia}
 \thanks{These authors contributed equally.}
 \affiliation{School of Chemistry, Chemical Engineering and Biotechnology, Nanyang Technological University, 62 Nanyang Drive, Singapore 637459}
\author{Ran Ni}
 \email{r.ni@ntu.edu.sg}
 \affiliation{School of Chemistry, Chemical Engineering and Biotechnology, Nanyang Technological University, 62 Nanyang Drive, Singapore 637459}


\date{\today}

\begin{abstract}
Materials incorporating covalent adaptive networks (CAN), e.g., vitrimers, have received significant scientific attention due to their distinctive attributes of self-healing and stimuli-responsive properties.
Different from direct crosslinked systems, bivalent and multivalent systems require a bond swap algorithm that respects detailed balance, considering the multiple equilibria in the system. 
Here we propose a simple and robust algorithm to handle bond swap in multivalent and multi-species CAN systems. 
By including a bias term in the acceptance of Monte Carlo moves, we eliminate the imbalance from the bond swap site selection and multivalency effects, ensuring the detailed balance for all species in the system.
\end{abstract}


\maketitle 

\section{introduction}
\label{sec:level1}

Covalent adaptive networks (CANs) or dynamic covalent networks (DCNs) in polymeric materials have been widely studied 
for their promising \rt{applications} in self-healing~\cite{chen2019dynamic, chen2019injectable}, stimuli-responsive~\cite{chen2019dynamic, haldar2015polyisobutylene}, and shape-memory\cite{yang2020assembling, chen2017multi} materials.
A novel CAN system, vitrimer, recently emerged, exhibiting great potential across material and biological domains~\cite{zheng2021vitrimers, krishnakumar2021vitrimers}.
Vitrimers not only exhibit superior mechanical properties likewise thermosets but also retain plasticity and reprocessability,
thanks to their exchangeable covalent crosslinking network~\cite{montarnal2011silica}.
The interlinking covalent bonds can swap between different polymer chains, leading to the dynamic topological change in the system.
Therefore, they behave like thermosets at low temperatures due to slow bond exchanges, but become viscoelastic at higher temperatures due to rapid bond swapping.

Studies of CAN systems call for algorithms to simulate dynamic bond swaps in coarse-grained models.
Beside the modified Kern-Frenkel model~\cite{smallenburg2013patchy, smallenburg2013liquids} and three-body potential~\cite{sciortino2017three}, 
hybrid Monte Carlo (MC) molecular dynamics methods based on the Metropolis-Hastings algorithm~\cite{wu2019dynamics, stukalin2013self} have been primarily used for simulating CAN systems.
Such algorithms propose potential bond swap sites randomly and execute swaps based on a preset probability, which is controlled by the bond swap rate. 
When generalizing those algorithms to multivalent, multi-species systems, such as linker-mediated vitrimers~\cite{rottger2017high,lei2020entropy,xia2022entropy}, 
it is essential to respect the detailed balance.
\rt{Beyond the MC algorithm commonly handling multivalent hybridization in DNA-coated colloids~\cite{leunissen2011numerical}, the algorithm handeling multivalent CAN systems requires to consider the combinatorial entropy in each MC move to respect the detailed balance~\cite{lei2020entropy,xia2022entropy}.}
Although one can use a three-body potential to ensure the detailed balance, to model bond swaps using an elaborated continuous three-body potential,
designing a proper three-body or multi-body potential for a multivalent systems remains challenging.
\rt{Additionally, the three-body potential method may misbehave at high density, where interactions between more than two particles are frequent.}
\rt{The continuous three-body potential may introduce an effective repulsion, which can affect the thermodynamics of the system.}

In this work, we propose an algorithm that can model bond swaps not only in monovalent and bivalent systems but also in multivalent systems while respecting the detailed balance~\cite{lei2020entropy,xia2022entropy}.
We verify our algorithm in both monovalent and multivalent ideal scenarios and demonstrate the accuracy and robustness of the algorithm.

\section{algorithm}

\begin{figure}[!ht]
    \includegraphics[width = 0.55\textwidth]{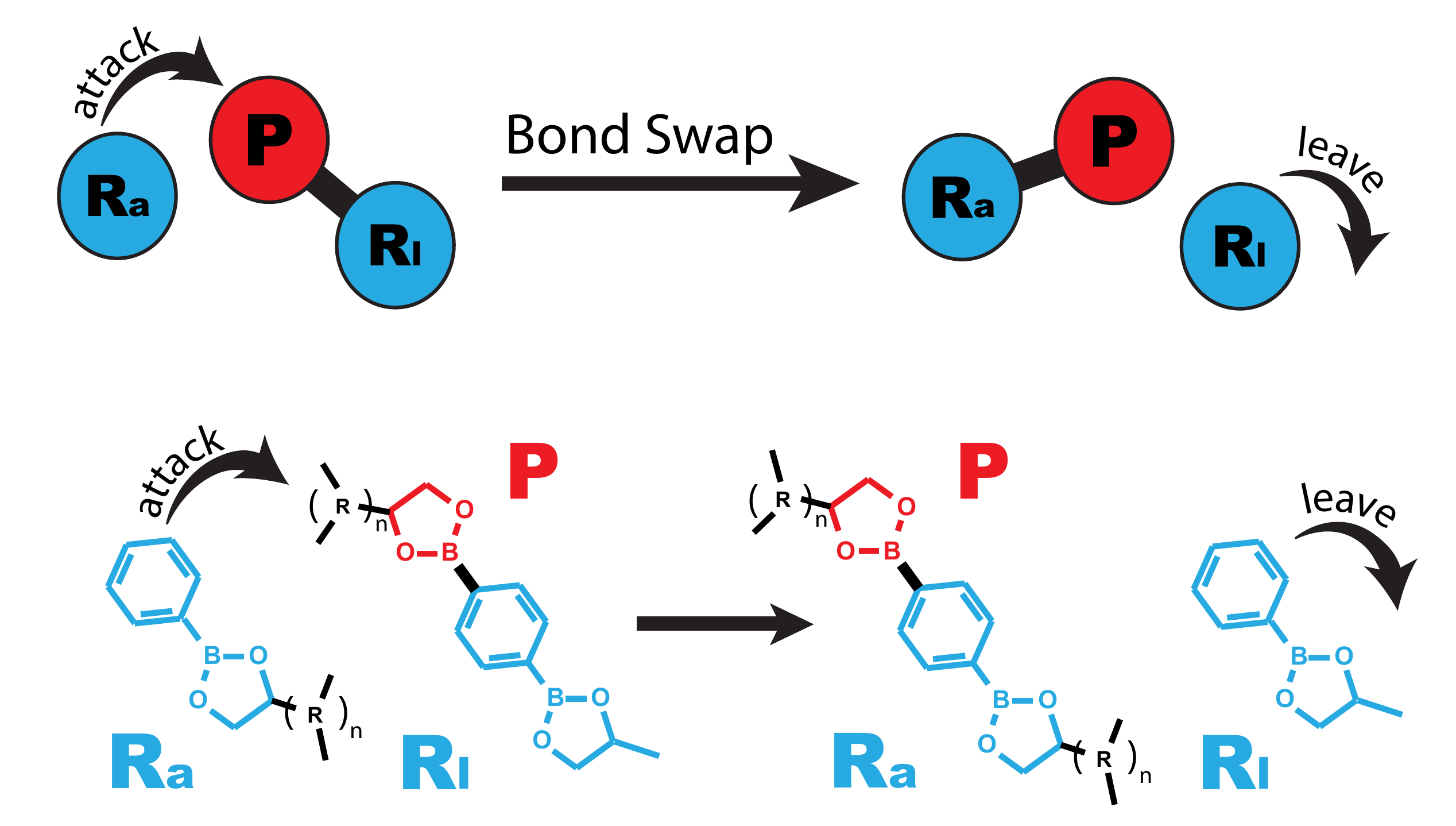}
    \caption{\label{fig:1}\footnotesize  \textbf{A general bond swap schematic.} Top: a pivot $\rm P$ is attacked by a residue ${\rm R}_a$ and forms a new bond, while breaking an existing bond and release a leaving residue ${\rm R}_l$. 
    Through the whole process, the bond is swapped from ${\rm R}_l$ to ${\rm R}_a$, and the total \rt{valences} are conserved.
    \rt{Bottom: a typical chemical realization (dioxaborolane vitrimer crosslinking~\cite{rottger2017high}) of the model.}}
\end{figure}

A typical bond swap coarse-grained model consists of two primary particle types: pivot species $\rm P$ and residue species $\rm R$.
These particles reversibly bond with each other in the system,
with each pivot maintaining a fixed valence $v_p$, and each residue varying its valence from $0$ to the maximum $v_r$.
As shown in Fig.~\ref{fig:1}, in a typical bond swap MC move, an attacking residue ${\rm R}_a$, with at least one unoccupied valence, forms a new bond with a pivot particle $\rm P$, 
meanwhile $\rm P$ breaks one bond with a leaving residue ${\rm R}_l$.
\rt{A typical chemical realization of this model (dioxaborolane vitrimer crosslinking~\cite{rottger2017high}) is shown in Fig.~\ref{fig:1}.}
In this process, ${\rm R}_a$ gains one valence while the ${\rm R}_l$ loses one, 
thus preserving the overall valences in the system.
A typical Metropolis-Hastings MC bond swap algorithm randomly proposes a potential bond swap pair $({\rm P}\text{-}{\rm R}_l, {\rm R}_a)$ in the system with a proposal probability $p$ 
and executes a bond swap trial move with an acceptance $acc$~\cite{wu2019dynamics,stukalin2013self}. 

\begin{figure*}[!ht]
    \includegraphics[width = \textwidth]{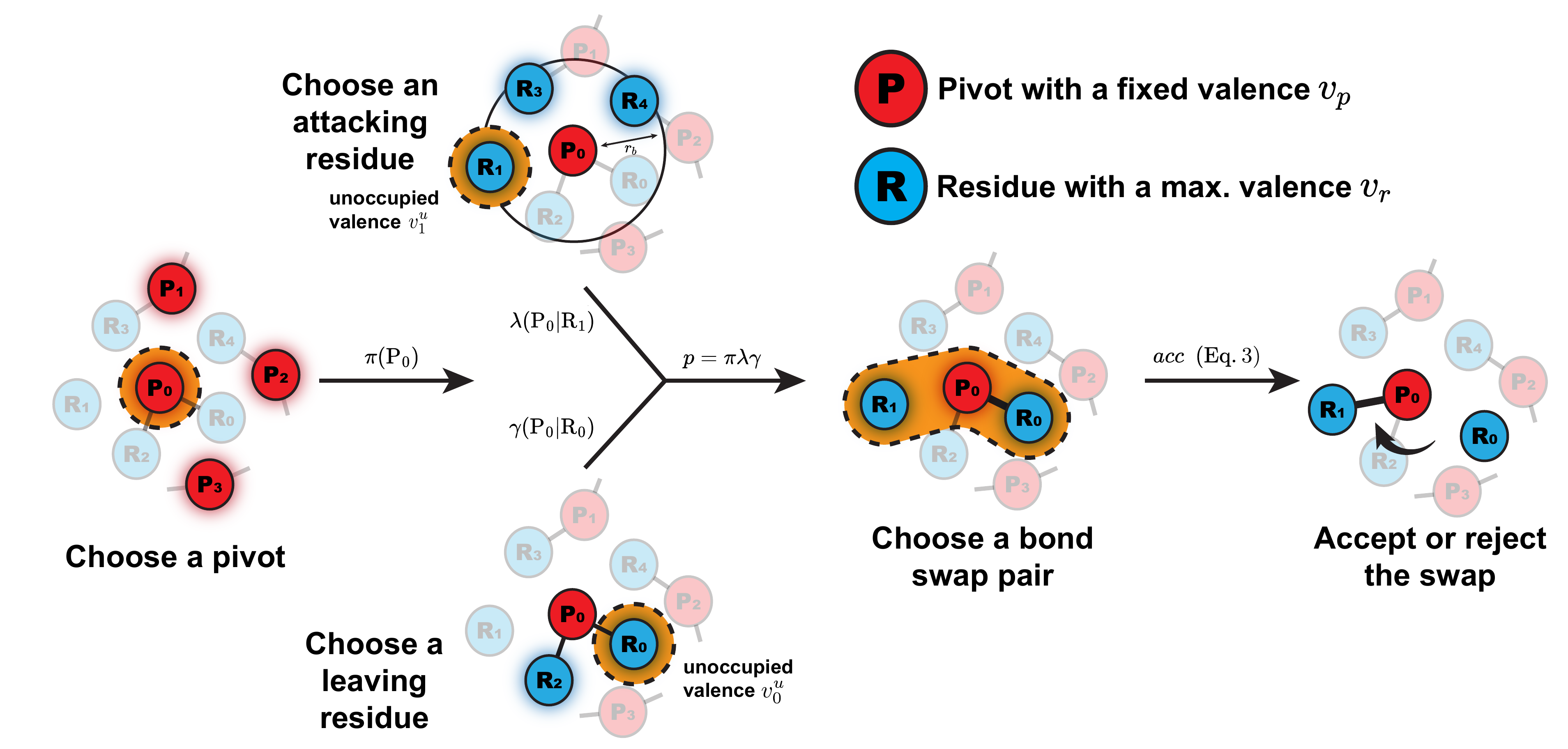}
    \caption{\label{fig:2} \footnotesize \textbf{Schematic of the bond swap algorithm.} 
    The algorithm first chooses a random pivot ($\rm P_0$) from the sampling space with the probability $\pi(\rm P_0)$.
    Then it searches for all residues with vacant \rt{valences} within $r_b$ and randomly selects one as the attacking residue ($\rm R_1$) with the probability $\lambda(\rm R_1|\rm P_0)$. 
    At the same time, it randomly chooses a leaving residue ($\rm R_0$) from $\rm P_0$ with the probability $\gamma(\rm R_0|\rm P_0)$.
    The probability $p$ to choose $(\rm P_0\text{-}\rm R_0, \rm R_1)$ is $\pi(\rm P_0)\lambda(\rm R_1|\rm P_0)\gamma(\rm R_2|\rm P_0)$, 
    and the acceptance can be calculated through Eq ~\ref{eq:3}.}
\end{figure*}

In a generic scheme as shown in Fig.~\ref{fig:2}, a random bond swap pair $({\rm P}_0 \text{-}{\rm R}_0, {\rm R}_1)$ is selected with a proposal probability $p$ using an ergodic stochastic algorithm.
Here, ${\rm R}_1$ is the attacking residue with unoccupied valences $v^u$, and ${\rm R}_0$ is the leaving residue.
Once chosen, the pair $(\rm P_0 \text{-}\rm R_0, \rm R_1)$ is given a specific acceptance $acc$ to react to form $(\rm P_0 \text{-}\rm R_1, \rm R_0)$.
For the corresponding inverse process, where now $\rm R_0$ is the attacking residue and $\rm R_1$ is the leaving residue with unoccupied valences $v'^u$, 
the pair $(\rm P_0 \text{-}\rm R_1, \rm R_0)$ is selected with \rt{a} reverse proposal probability $p'$.
This pair is given a specific reverse acceptance $acc'$ to react back to form $(\rm P_0 \text{-}\rm R_0, \rm R_1)$.
The detailed balance condition in the scheme implies~\cite{frenkel2001understanding}:
\begin{equation} \label{eq:1}
    \frac{p/v^u \ acc}{p'/v'^u \ acc'} = \exp(-\beta\Delta G_0), 
\end{equation}
where $\Delta G_0$ is the reaction free energy change; $\beta = 1/k_B T $  with $k_B$ and $T$ the Boltzmann constant and the temperature of the system, respectively. 
Here, we divide $p$ and $p'$ by the total unoccupied valences of the attacking residue $v^u$ and $v'^u$, respectively, as we consider each unoccupied valence having the same probability to react with the pivot.
Most existing bond swap algorithms assume the bond swap pair proposal process is symmetric, implying \rt{$(p/(v^u))/(p'/(v'^u)) = 1$}~\cite{wu2019dynamics, stukalin2013self}.
However, this may not be true in multivalent and multi-species systems.
Therefore, we cannot simply have $acc/acc' = \exp(-\beta\Delta G_0)$.
To comply with Eq. 1, we introduce a bias term to the acceptance of each MC move, and according to the Metropolis-Hastings rule~\cite{metro1953,Hastings}, the acceptance of an MC move is	
\begin{equation}
    acc = \min \left( 1,\ \frac{p'/v'^u}{p/v^u} \exp(-\beta\Delta G_0) \right). \label{eq:2}
\end{equation}

\subsection{Bond Swap Pair Proposal Algorithm}

It is important to note that any ergodic bond swap pair proposal algorithm can be employed in the MC move, provided the proposal probabilities $p$ and $p'$ are known.
Here, we propose a symmetric proposal algorithm with $p = p'$ to \rt{simplify} the acceptance in Eq.~\ref{eq:2},
and the schematic of the algorithm is illustrated in Fig \ref{fig:2}.
A covalent bond is treated as an infinitely deep square well potential with the bond length $r_{b}$~\cite{lei2020entropy,xia2022entropy},
and a bond swap move can occur only if the center-to-center distance between a pivot and an attacking residue is less than $r_{b}$.
In each bond swap pair proposal, the algorithm first chooses a random pivot $\rm P_0$ from all pivots in the system with the probability $\pi(\rm P_0)$.
It then searches for all potential attacking residue candidates with any unoccupied valence near $\rm P_0$ within $r_{b}$,  
and randomly selects an attacking residue $\rm R_1$ with unoccupied valences $v_1^u$ with the probability $\lambda(\rm R_1|\rm P_0)$.
Lastly, it selects a leaving residue $\rm R_0$ with unoccupied valence $v_0^u$ with the probability $\gamma(\rm R_0|\rm P_0)$ from $\rm P_0$.
Similarly, $\pi'(\rm P_0)$, $\lambda'(\rm R_0|\rm P_0)$, and $\gamma'(\rm R_1|\rm P_0)$ are the probabilities of inverse proposal,
with the unoccupied valences $v'^u$ of the inverse attacking residue $\rm R_0$, which is equal to $v_0^u + 1$. 
Given that the total pivot number $N_{\rm P}$ in the system, the total number of potential attacking residue candidates $N_{att}$, and the total number of potential leaving residue candidates  $N_{lv}$ are the same for both forward and inverse proposals, 
$\pi({\rm P_0}) = \pi'({\rm P_0}) = 1/N_{\rm P}$, $\lambda({\rm R_0|\rm P_0}) = \lambda'({\rm R_1| \rm P_0}) = 1/N_{att}$, and $\gamma({\rm R_1|\rm P_0}) = \gamma'({\rm R_0| \rm P_0}) = 1/N_{lv}$. 
Therefore, Eq. \ref{eq:2} can be rewritten as:
\begin{equation}
    acc = \min \left(1,\ \frac{v_1^u}{v_0^u + 1} \exp(-\beta\Delta G_0) \right). \label{eq:3}
\end{equation}

To offer an overview of our algorithm, a pseudo-code is shown in Alg \ref{alg:1}.
\begin{algorithm}[!htbp]
    \DontPrintSemicolon
    \SetKwFunction{SwapPair}{SwapPair}
    \SetKwProg{Function}{Function}{:}{end}

    \caption[]{Bond swap Monte Carlo for coarse-grained models} \label{alg:1}
    \KwData{A coarse-grained bond swap system, 
    which contains pivot particles $\rm P$ with fixed valences $v_p$ and residue particles $R$ with maximal valences $v_r$. 
    The bond length is $r_b$}
    \KwData{$\beta$: thermodynamic beta $1/k_BT$, $k_B$ Boltzmann constant, $T$ temperature}
    \KwData{\SwapPair(): return a random bond swap pair with residue unoccupied valences and reaction free energy}

    \BlankLine
    
        Configuration initialization\;
        \While(){Sampling continuing}{
            $\{({\rm P}_0\text{-}{\rm R}_0, {\rm R}_1), v_0^u, v_1^u\} \gets$ \SwapPair()\;
            $acc \gets \min \left(1,\ \frac{v_1^u}{v_0^u + 1} \exp(-\beta\Delta G_0) \right)$ (Eq \ref{eq:3})\;
            \eIf(){a random $u \in [0, 1] \leq acc$}{
                Accept the bond swap MC move \;
            }(){
                Reject the bond swap MC move \;
            }
        }

    \BlankLine
    
        \Function{\SwapPair{}}{
            Randomly choose a \textbf{pivot} $\rm P_0$ among all pivots in the system.\;
            Search all attacking residue candidates with unoccupied valences near $\rm P_0$ within the bond length $r_b$, 
            and randomly choose a residue $\rm R_1$ with unoccupied valences $v_1^u$ as the \textbf{attacking residue}\;
            Randomly choose a residue $\rm R_0$ with unoccupied valences $v_0^u$ bonded with $\rm P_0$ as the \textbf{leaving residue}\;
            \KwRet $\{({\rm P}_0\text{-}{\rm R}_0, {\rm R}_1), v_0^u, v_1^u\}$\;
        }

\end{algorithm}

\section{Algorithm validation}
In the following, we verify our algorithm in various systems including an ideal monovalent diatomic system, an ideal multivalent linker system and an ideal binary chain system, in which we compare the simulation results with theoretical predictions.

\subsection{Ideal monovalent diatomic system}

\begin{figure}[!ht]
    \includegraphics[width = 0.6\textwidth]{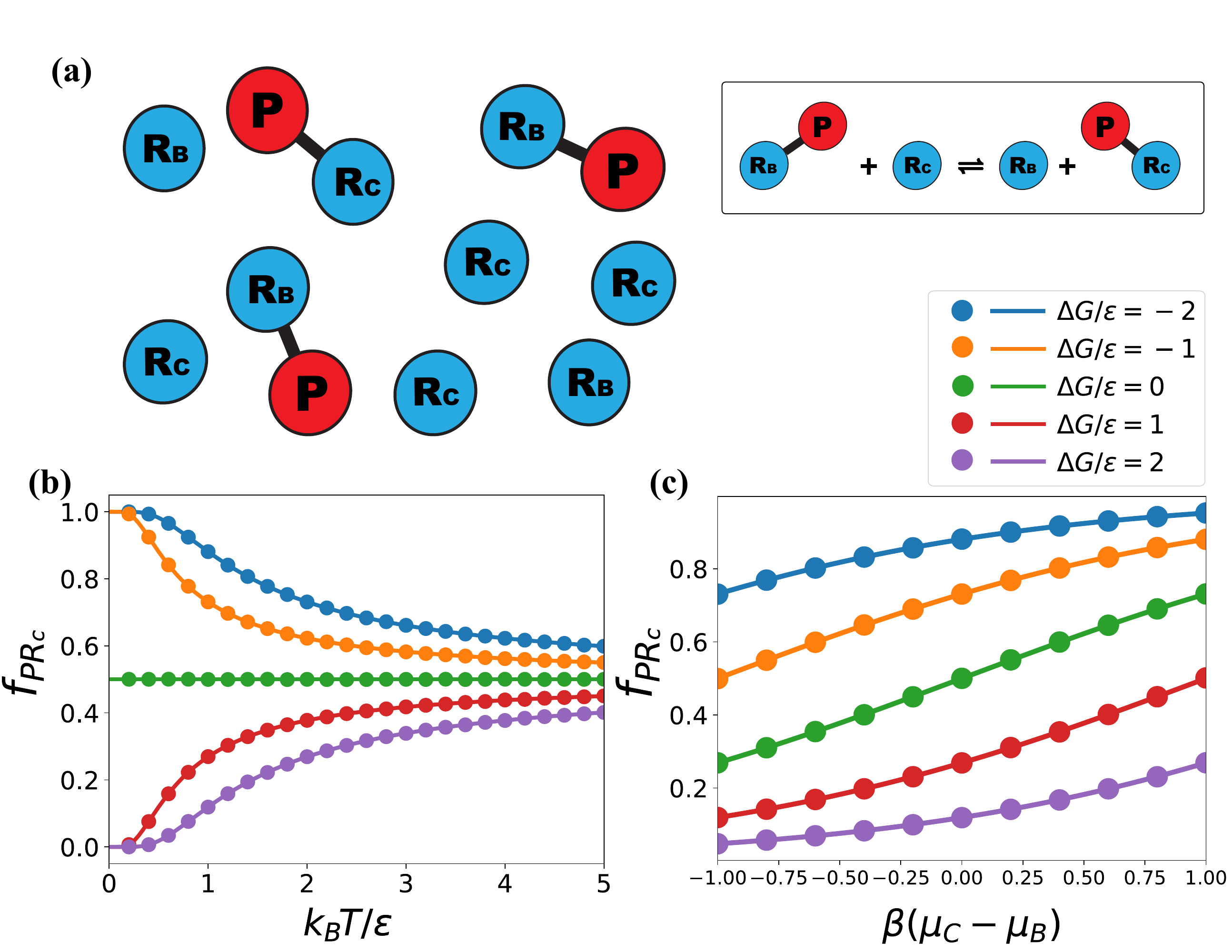}
    \caption{\label{fig:3} \footnotesize \textbf{Monovalent diatomic system.} 
    (a) Schematic of \rt{a} monovalent diatomic system, incorporating one species of pivot $\rm P$ and two species of residues $\rm R_B$, $\rm R_C$. 
    (b) \& (c), $f_{\rm PR_C}$ as the functions of temperature $k_B T/\epsilon$ 
    and the chemical potential difference of $\rm R_C$ and $\rm R_B$, $\mu_C - \mu_B$,
    where $\rm P$ is the pivot with a fixed valence $v_p = 1$, $\rm R_B$ and $\rm R_C$ are residues with a maximal valence $v_r = 1$. 
    The solid curves represent the theoretical predictions in Eq.~\ref{eq:mono}, and the symbols are the simulation results.
    In all simulation and theoretical results, except their independent variables, $\mu_{\rm B} = \mu_{\rm C} = -0.916$, $N_{\rm P} = 400$, $\Delta G / \epsilon = -1$.
    The size of the simulation box is $20r_b$. Each simulation requires $5\times 10^5$ MC moves to equilibrate and $1\times 10^8$ MC moves to sample.}
\end{figure}

We first consider a system of ideal monovalent diatomic molecules, in which $v_p = 1$ and $v_r = 1$. 
The system consists of one type of pivot $\rm P$ with $v_p = 1$ and two other types of residues $\rm R_B$, $\rm R_C$ with $v_r = 1$.
Initially, each pivot $\rm P$ is bonded with either a $\rm R_B$ or $\rm R_C$, 
and these bonds can interchange through the algorithm above as shown in Fig.~\ref{fig:3}a. 
\rt{Given that there are just two kinds of residues, the only reaction in the system is}:
\begin{equation} \label{eq:mono_eq}
    \rm{PR_B} + \rm{R_C} \rightleftharpoons \rm{PR_C} + \rm{R_B}.
\end{equation}
Besides the bonding interactions, the interaction between all particles is modelled as an ideal gas, and the equilibrium state of the system can be described via the chemical equilibrium relationship:
\begin{equation} \label{eq:mono}
    \frac{f_{\rm PR_C}}{f_{\rm PR_B}} = \exp{\left(-\beta \Delta G + \beta \mu_{\rm C} - \beta \mu_{ \rm B}\right)},
\end{equation}
where $f_{\rm PR_B}$, $f_{\rm PR_C}$ are the fractions of $\rm P$ in those molecules over all $\rm P$ particles,
and $\mu_{\rm B}$, and $\mu_{\rm C}$ are the chemical potentials of species $\rm{R_B}$ and $\rm{R_C}$, respectively.
Substituting Eq. \ref{eq:mono} into $f_{\rm PR_B}+f_{\rm PR_C} = 1$ yields the equilibrium fraction of each component.
To compare with the theoretical prediction, we perform grand canonical Monte-Carlo (GCMC) simulations,  
and the simulation results are presented in Fig. \ref{fig:3}. 
The number of pivot particles is fixed at $N_{\rm P} = 400$.
The obtained fraction of $\rm P$ in $f_{\rm P R_C}$ in systems at different temperature $k_B T/\epsilon$ of various chemical potential difference $\mu_{\rm C} - \mu_{\rm B}$ from computer simulations (symbols) in comparison with theoretical predictions (solid curves) are shown in Fig.~\ref{fig:3}b and c, in which one can find an excellent agreement. Here $\epsilon$ is the energy unit in the system.

\subsection{Ideal multivalent linker system}

\begin{figure}[!ht]
    \includegraphics[width = 0.6\textwidth]{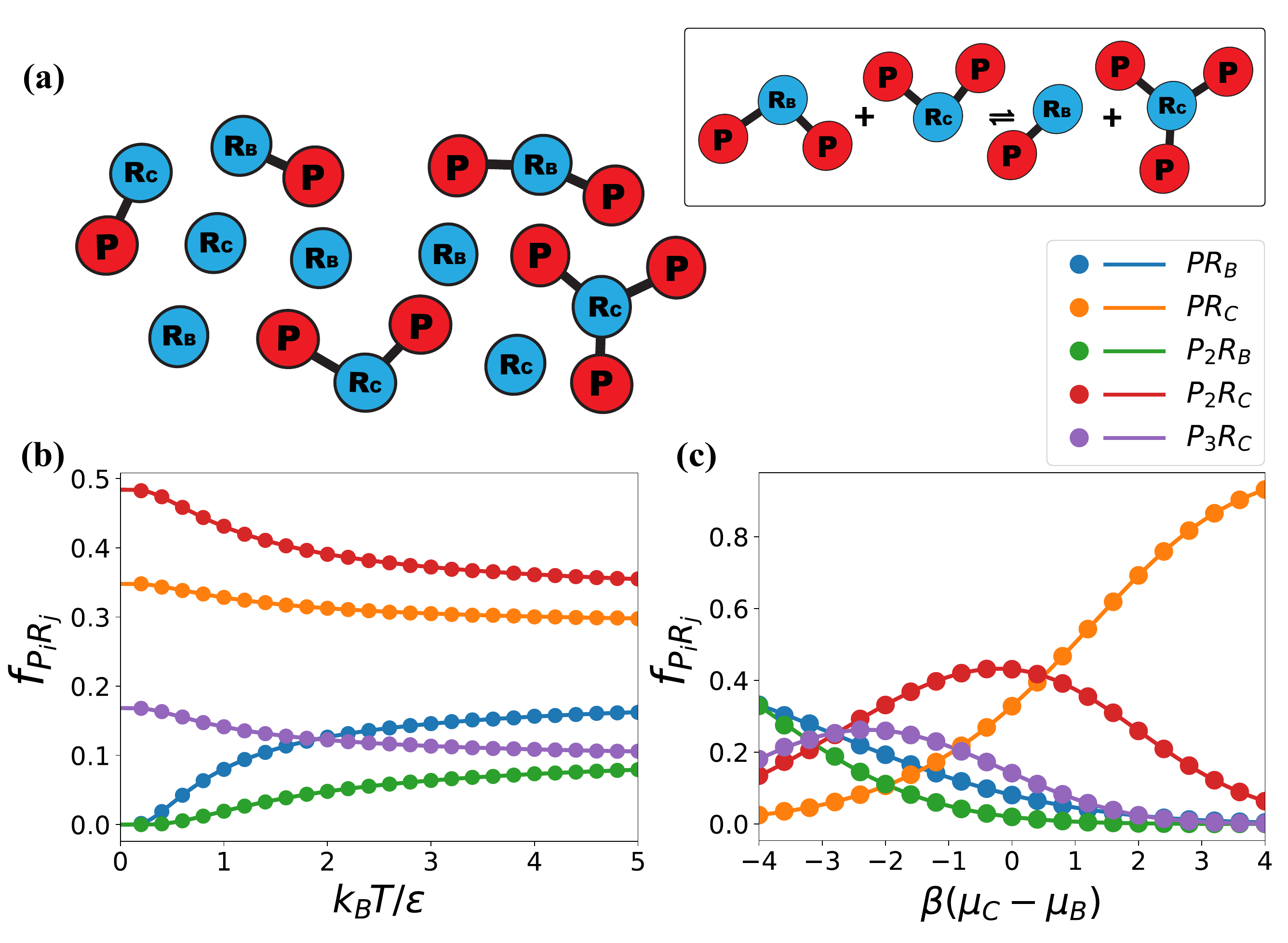}
    \caption{\label{fig:4} \footnotesize \textbf{Multivalent linker system} 
    (a), Schematic of an multivalent linker system, incorporating one species of pivot $\rm P$ and two species of residues $\rm R_B$, $\rm R_C$, 
    with polyatomic molecules $\rm P_2R_B$, $\rm P_2R_C$ and $\rm P_3R_C$.
    (b) \& (c) Pivot particle fraction in each pivot-residue molecule $f_{\rm PR_B}$, $f_{\rm PR_C}$,  T $f_{\rm P_2R_B}$, $f_{\rm P_2R_C}$, and $f_{\rm P_3R_C}$ as functions of temperature $k_B/\epsilon$, 
    the chemical potential difference of $\rm R_C$ and $\rm R_B$, $\mu_C - \mu_B$,
    where $\rm P$ is the pivot with a fixed valence $v_p = 1$, $\rm R_B$ and $\rm R_C$ are residues with maximal valence $v_{r_B} = 2$, $v_{r_C} = 3$. 
    The solid curves represent the theoretical predictions in Eq.~\ref{eq:multi}, and the symbols are the simulation results.
    In all simulation and theoretical results, except their independent variables, $\mu_{\rm B} = \mu_{\rm C} = -2.133$, $N_{\rm P} = 2400$, $\Delta G /\epsilon = -1$.
    The size of the simulation box is $30r_b$. Each simulation requires $5\times 10^5$ MC moves to equilibrate and $1\times 10^8$ MC moves to sample.}
\end{figure}

Next, we test the bond swap algorithm in systems of multivalent linkers, in which $v_p = 1$ but $v_r > 1$, and we consider no interaction between the linkers except bonding. 
As shown in Fig.~\ref{fig:4}a, star molecules ${\rm P}_i {\rm R_B}$ and ${\rm P}_j {\rm R_C}$, where $i, j < v_r$, can form in the system.
In each reaction, a residue particle with any unoccupied valance can seize a pivot particle from another residue particle, and the general reaction can be written as,
\begin{equation} \label{eq:multi_eq}
    {\rm P}_i {\rm R_B} + {\rm P}_j {\rm R_C} \rightleftharpoons {\rm P}_{i-1} { \rm R_B} + {\rm P}_{j+1} {\rm R_C}.
\end{equation}
Under the ideal condition, the equilibrium fraction of those species can be represented as,
\begin{equation} \label{eq:multi}
    \begin{aligned}
        \frac{f_{{\rm P}_i {\rm R_B}}}{f_{{\rm PR_B}}^i} =& \frac{\binom{v_{r_{\rm B}}}{i}}{\binom{v_{r_{\rm B}}}{1}^i}\exp{\left(-(i-1)\beta\mu_{\rm B}\right)},\\
        \frac{f_{{\rm P}_j {\rm R_C}}}{f_{{\rm PR_B}}^j} =& \frac{\binom{v_{r_{\rm C}}}{i}}{\binom{v_{r_{\rm C}}}{1}^i}\exp{\left(-j\beta \Delta G + \beta \mu_{\rm C} - j\beta \mu_{\rm B}\right)},
    \end{aligned}
\end{equation}
where $f_{{\rm P}_i {\rm R_B}}$ and $f_{{\rm P}_j {\rm R_C}}$ are the fraction of species over all $\rm P$ particles, and $v_{r_{\rm B}}$, $v_{r_{\rm C}}$ are the maximum valences of $\rm R_B$, $\rm R_C$, with $i \le v_{r_{\rm B}}$, $j \le v_{r_{\rm C}}$.
Terms $\binom{v_{r_{\rm B}}}{1}$, $\binom{v_{r_{\rm B}}}{i}$ and $\binom{v_{r_{\rm B}}}{i}$ are binomial coefficents describing the possible combinations of molecules $\rm PR_B$, ${\rm P}_i {\rm R_B}$ and ${\rm P}_j \rm R_C$, respectively.
Substituting Eq. \ref{eq:multi} into $\sum_{i = 1}^{v_{r_{\rm B}}}{f_{{\rm P}_i {\rm R_B}}} + \sum_{j = 1}^{v_{r_{\rm C}}}{f_{{\rm P}_j {\rm R_C}}} = 1$, 
yields the equilibrium fraction of each component by solving a polynomial equation.
To compare with the theoretical prediction, we perform GCMC simulations for systems consisting of one type of pivot $\rm P$ ($v_p = 1$) and two types of residues of $\rm R_B$ ($v_{r_{\rm B}} = 2$) and $\rm R_{\rm C}$ ($v_{r_{\rm C}} = 3$) at various temperature $k_B T/\epsilon$ and chemical potential differences $\mu_{\rm C} - \mu_{\rm B}$, of which the results are plotted in Fig \ref{fig:4}. 
One can see that the theoretical prediction (solid curves, Eq.~\ref{eq:multi}) agrees quantitatively with the fractions of different species obtained in computer simulations (symbols).
This verifies our bond swap algorithm in determining the equilibrium fractions of reactants in systems with complex reaction equilibria. 

\subsection{Ideal binary chain system}

\begin{figure}[!ht]
    \includegraphics[width = 0.65\textwidth]{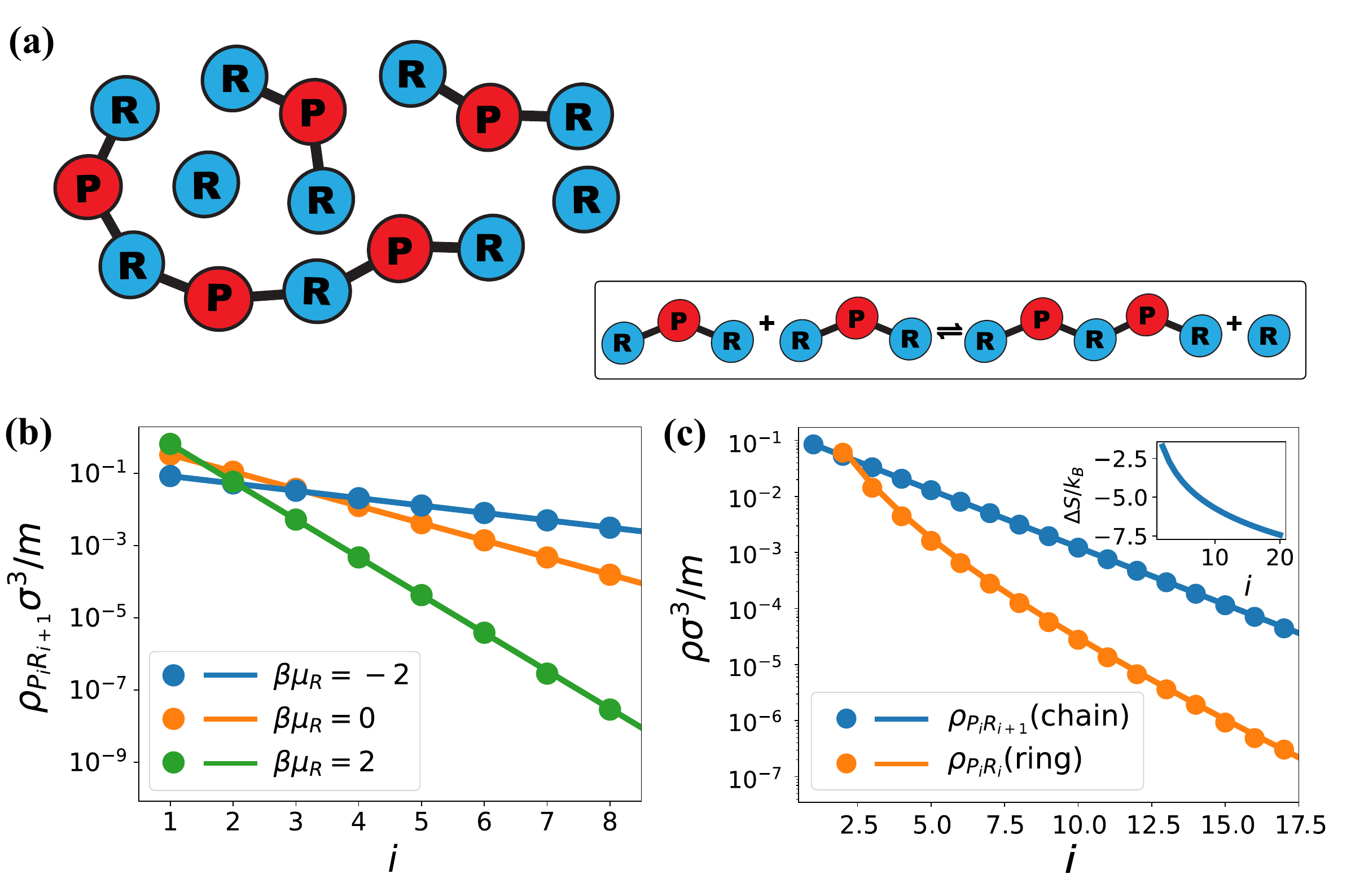}
    \caption{\label{fig:5} \footnotesize \textbf{Binary chain system} 
    (a) Schematic of a binary chain system, incorporating one species of pivot $\rm P$ and one species of residues $\rm R$, 
    with binary AB type long chain ${\rm P}_i {\rm R}_{i+1}$.
    (b)  Densities of binary chains of different length $\rho_{{\rm P}_i {\rm R}_{i+1}}$ as the functions of chain length $i$ (the number of pivots per chain) at different residue chemical potentials $\beta\mu_{\rm R}$,
    \rt{and (c) densities of binary chains $\rho_{{\rm P}_i {\rm R}_{i+1}}$ and rings $\rho_{{\rm P}_i {\rm R}_i}$ of different length as the functions of chain length $i$ when $\beta\mu_R = -2$,}
    where $\rm P$ is the pivot with a fixed valence $v_p = 2$, $\rm R$ is residue with maximal valence $v_r = 2$. 
    The solid curves represent the theoretical predictions in Eq.~\ref{eq:chain} and Eq.~\ref{eq:ring2}, and the symbols are the simulation results.
    \rt{(inset) Theoretical predictions of total entropy loss (configurational and combinatorial) $\Delta S/k_B$ of chains forming rings with increasing chain length $i$.}
    In all simulation and theoretical results, except their independent variables, $T = 1$, $N_{\rm P} = 800$, $\Delta G = 0$.
    The size of the simulation box is $20r_b$. Each simulation requires $5\times 10^5$ MC moves to equilibrate and \rt{$5\times 10^8$} MC moves to sample.}
\end{figure}

Lastly, we test our bond swap algorithm in a general system where $v_p > 1$ and $v_r > 1$.
Pivots and residues can form large clusters in the system since both of them are multivalent.
For simplicity, we consider the system, in which $v_p = 2$ and $v_r = 2$.
The pivots and residues can form AB type long chains as illustrated in Fig.~\ref{fig:5}a.
The general chain elongation reactions are
\begin{equation} \label{eq:chain_eq}
    {\rm P}_i {\rm R}_{i+1} + {\rm P}_j {\rm R}_{j+1} \rightleftharpoons {\rm P}_{i+j} {\rm R}_{i+j+1} + \rm{R}.
\end{equation}
The system is similar to the polymer linear condensation problem, where the density of $i$-mers follows a geometric series~\cite{rubinsten2003polymer},
and the densities of different long chains can be represented as,
\begin{equation} \label{eq:chain}
    \frac{\rho_{{\rm P}_i {\rm R}_{i+1}}}{\rho_{{\rm PR}_2}^i} = \exp{\left(-(i-1)\beta\mu_R\right)},
\end{equation}
where $\rho_{{\rm P}_i {\rm R}_{i+1}}$ and $\rho_{{\rm PR}_2}$ are the densities of long chains ${\rm P}_i {\rm R}_{i+1}$ and \rt{monomer}, respectively, and $\mu_R$ represents the chemical potential of residue $\rm R$.
We substitute the $\rho_{{\rm PR}_2}$ obtained from the simulation in Eq.~\ref{eq:chain} to obtain the densities of longer chains.
Accordingly, the slope of logarithm density versus chain length is ${\rm d} \ln (\rho_{{\rm P}_i {\rm R}_{i+1}}) / {\rm d} i = \ln ( \rho_{\rm {PR}_2})-\beta\mu_{ R}$ and decreases with increasing $\mu_{ R}$.
\rt{Furthermore, rings may form in the system with the general reaction:}
\begin{equation} \label{eq:ring_eq}
    \rt{{\rm P}_i {\rm R}_{i+1} ({\rm chain}) \rightleftharpoons {\rm P}_{j} {\rm R}_{j} ({\rm ring}) + {\rm P}_{i-j}{\rm R}_{i-j+1} ({\rm chain}).}
\end{equation} 
\rt{Different from Eq.~\ref{eq:chain}, the ring formation causes both configurational and combinatorial entropy loss, and $\rho_{{\rm P}_i {\rm R}_{i}}$, the density of ring ${\rm P}_i{\rm R}_i $, can be written as}
\begin{equation} \label{eq:ring}
    \rt{\frac{\rho_{{\rm P}_i {\rm R}_{i}}}{\rho_{{\rm PR}_2}^i} = \exp{\left(-i\beta\mu_R + \left(\Delta S_{\rm{conf}}^i + \Delta S_{\rm{comb}}^i\right)/k_B\right)}\quad(i > 1),}
\end{equation} 
\rt{where $\Delta S_{\rm{conf}}^i$ and $\Delta S_{\rm{comb}}^i$ are the configurational and combinatorial entropy loss of a ring ${\rm P}_i {\rm R}_{i}$ forming from a chain ${\rm P}_i {\rm R}_{i}$.}
\rt{$\Delta S_{\rm{conf}}^i$ can be calculated by the configurational free volume of chains and rings, which can be obtained by integrating the end-to-end distance distribution of the chain ${\rm P}_i {\rm R}_{i}$~\cite{rubinsten2003polymer}:}
\begin{eqnarray} \label{eq:Sconf1}
    \rt{\Delta S_{\rm{conf}}^i/k_B = \ln\left(\frac{\Omega_i^{\text{ring}}}{\Omega_i^{\text{chain}}}\right)} &&\rt{= \ln\left(\int_{0}^{r_b}{D_{2i-1}(r)4\pi r^2dr}\right)} \\
    &&\rt{= \ln\left(\int_{0}^{r_b}{(\underbrace{D_1 \ast D_1 \ast \cdots \ast D_1}_{2i-1})(r)4\pi r^2dr}\right),}
\end{eqnarray}
\rt{where $\Omega_i^{\text{ring}}$ and $\Omega_i^{\text{chain}}$ are the configurational free volume of a ring and a chain of $i$ length, respectively.}
\rt{$D_N(r)$ is the probability density function of end-to-end distance distribution of $N$-chain,}
\rt{and is exactly the $N$-time convolution of $D_1(r)$ which is a uniform distribution in a sphere with a variance $\sigma^2 = 5r_b^2$.}
\rt{The multiple convolution can be estimated by central limit theorem, and the integral becomes $\chi_3$ distribution form (Maxwell distribution).}
\begin{eqnarray} \label{eq:Sconf2}
    \rt{\Delta S_{\rm{conf}}^i/k_B} &&\rt{\approx \ln\left(\int_{0}^{r_b}{f_{\chi_3}\left(\frac{r}{\sqrt{2i-1}\sigma}\right)/(\sqrt{2i-1}\sigma)}dr\right)} \\
    &&\rt{\approx \ln\left(P\left(\frac{3}{2}, \frac{5}{4i-2}\right)\right),}
\end{eqnarray} 
\rt{where $f_{\chi_3}(x)$ is the probability density function of $\chi_3$ distribution, $P(s,x)$ is the regularized gamma function originating from the cumulative distribution function of $\chi_3$ distribution.}
\rt{As the permutation number of a $n$ particle ring is $2/n$ of that of a $n$ particle chain due to the ring symmetry, and }
\begin{equation} \label{eq:Scomb}
    \rt{\Delta S_{\rm{comb}}^i/k_B = \ln\left(1/i\right).}
\end{equation}
\rt{Therefore, Eq.~\ref{eq:ring} can be rewritten to}
\begin{equation} \label{eq:ring2}
    \rt{\frac{\rho_{{\rm P}_i {\rm R}_{i}}}{\rho_{{\rm PR}_2}^i} = \exp{\left(-i\beta\mu_R\right)P\left(\frac{3}{2}, \frac{5}{4i-2}\right) \frac{1}{i}.}}
\end{equation}
\rt{With increasing chain length $i$, the total entropy change $\Delta S = \Delta S_{\rm{conf}}^i + \Delta S_{\rm{comb}}^i$ of ring forming declines (Fig.\ref{fig:5}c inset).}
\rt{Moreover, $P(\frac{3}{2}, \frac{5}{4i-2})$ is asymptotic to $\frac{4}{3\sqrt{\pi}}(\frac{5}{4i-2})^{3/2}$ at large $i$,}
\rt{and the slope of the logarithm ring density versus $i$ is thus approximately ${\rm d} \ln (\rho_{{\rm P}_i {\rm R}_i}) / {\rm d} i \approx \ln(\rho_{{\rm PR}_2}) - \beta\mu_R - 3/(2i-1) - 1/i$ when $i$ is large,}
\rt{which is smaller than that of the chain density.}
\rt{As the result, the ring density declines faster than the chain density with increasing $i$.} 
We plot the theoretical predictions (Eq.~\ref{eq:chain} and Eq.~\ref{eq:ring}) in comparison with results obtained in GCMC simulations \rt{of chain densities} at three different $\mu_{\rm R}$ in Fig.~\ref{fig:5}b
\rt{and chain and ring densities at $\beta \mu_{ R} = -2$ in Fig.\ref{fig:5}c.}
\rt{Despite small discrepancy between ring density predictions and simulation results due to the approximation of $\Delta S_{\rm{conf}}^i$,}
one can see that the theoretical prediction (solid curves) agrees quantitatively with GCMC simulations (symbols).
This proves that our algorithm is able to handle polymerization or aggregation problems for various polymeric and biological systems.

\section{Discussion and Conclusion}

In this paper, we have developed a simple and robust algorithm for simulating dynamic bond swapping in multivalent and multi-species systems.
The algorithm respects the detailed balance through introducing a bias term in the acceptance of MC move.
Moreover, we provide universal guidelines for determining the bias term for any designed algorithm. 
Through calculating the bias term, our algorithm can be tailored to simulate any bond swapping systems, with the detailed balance.
\rt{It is worth mentioning that when the rigidity of the bond increases, the efficiency of the bondswap algorithm decreases, which can be resolved by using biased MC methods~\cite{martinez2010anomalous}.}
\rt{To simulate large and dense systems, the algortihm can be accelerated and adapted to most MC parallel schemes~\cite{thompson2022lammps, kampmann2015parallelized} with ease, since the bond swap move are as normal as other local MC moves.}
\rt{Nevertheless, one must guarantee the detailed balance during the parallelization and customize the bias term if necessary.}
Additionally, our algorithm can be also implemented in molecular dynamics (MD) by performing bond swap MC moves at a random time interval in MD simulations.
In such hybrid-MC-MD simulation, an activation energy barrier $\Delta G_a$ can be introduced to control the kinetics of bond swaps, and 
\begin{equation}
    acc = \exp(-\beta\Delta G_a)\min \left(1,\ \frac{p'/v'^u}{p/v^u} \exp(-\beta\Delta G_0) \right) \label{eq:MD}
\end{equation}
A higher $\Delta G_a$ implies a lower bond swap rate, and vice versa.
\rt{Not like the three-body potential method which may misbehave at high density, the hybrid-MC-MD algorithm does not introduce any artificial potential in MD simulations~\cite{lei2020entropy,xia2022entropy},}
\rt{and thus may get more accurate representation of the structural information.}
By tuning the activation energy barrier $\Delta G_a$, our algorithm can efficiently capture the kinetic information of dynamic crosslinking systems~\cite{xia2022entropy}.
Beyond these applications, the algorithm holds promise for simulating more complex multivalent soft matter systems, including DNA-coated colloids~\cite{xia2020linker,angioletti2014mobile}, nanocrystal assembly~\cite{kang2022colorimetric, lin2016linker} and some biological systems~\cite{harrington2010iron,holten2011ph,gordon2015dynamic}.

\section{Data Availability}

The data that support the findings of this study are available within the article and from the corresponding author upon reasonable request.


%
%

%

\begin{acknowledgments}
This work is supported by the Academic Research Fund from the Singapore Ministry of Education Tier 1 Gant (RG59/21), and the National Research Foundation, Singapore, under its 29th Competitive Research Programme (CRP) Call, (Award ID NRF-CRP29-2022-0002). 
\end{acknowledgments}

\nocite{*}

\bibliography{main}

\end{document}